# FREQUENCY DEPENDENCE OF SUBSURFACE FLAWS DETECTION EFFICIENCY VIA THE TOTAL INTERNAL REFLECTION ULTRASONIC SENSOR

## Yuliia Kominko*

* Yuliia Kominko, Taras Shevchenko National University of Kyiv, 64/13, Volodymyrska Street, City of Kyiv, Ukraine, 01601, e-mail: yuliia.kominko@gmail.com

*Here is proposed a method for detecting subsurface flaws in isotropic specimens via using a novel device The Total Internal Reflection Ultrasonic Sensor (TIRUS). Main features of TIRUS are the following: (i) the sensor is able to detect subsurface flaws; (ii) the detection depth is about the ultrasonic wavelength.*

### Introduction

The Total Internal Reflection Ultrasonic Sensor (TIRUS) is a device specially aimed at detecting subsurface flaws in a tested specimen based on the frustrated total internal reflection of bulk ultrasonic waves in the sensor body [1]. In the fabricated experimental sample of the TIRUS a transducer of the sensor emitted a slow shear wave with the velocity ~600 m/s (SSW) into the [110] direction of the TeO2 crystal (Fig. 1) [2]. The sensor design was such that the displacement vector of the SSW lay in the planes of reflecting surfaces of the crystal, that is, it was a shear horizontal (SH) wave with respect to these planes. It was assumed that with such a design no other waves should exist in the crystal except those polarized as the SSW [3]. Total internal reflection of a probing ultrasonic wave incident at the interface between the sensor and a tested object is frustrated if there is a subsurface flaw in the evanescent field under the interface (Fig. 1, a). As a result, the reflected wave is affected and the sensor output changes.

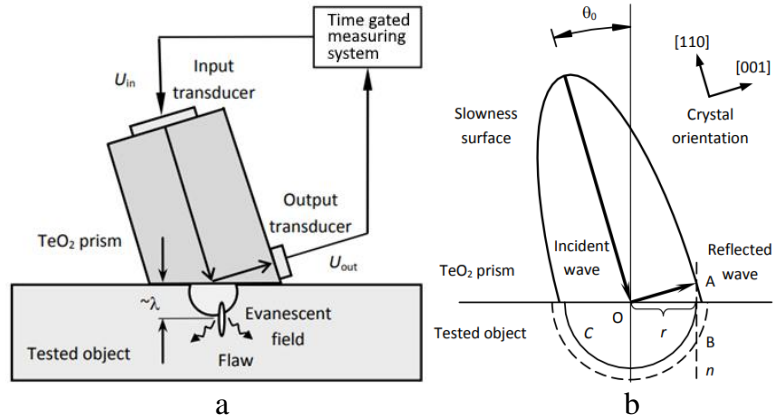

Fig. 1. The TIRUS concept. a – configuration of the sensor-tested object "assembly"; b – geometry of the total internal reflection of the probing slow shear wave in the TeO2 prism.

### Experiments of testing specimens with the TIRUS

An operational characteristic of the TIRUS can be defined as a two-port network`s transfer function $T(f) = \frac{U_{out}(f)}{U_{in}(f)} = |T(f)|\exp(i\varphi(f))$. In the case of a pulse input of the TIRUS, "time gated" insertion loss $IL_{TG}(f)$ and "time-gated" phase $\varphi_{TG}$ of the transfer function are:

$$IL_{TG} = 20\log U_{carr}/U_{in}, \qquad \varphi_{TG} = \varphi_{carr} - \varphi_{ref}$$

Here $U_{carr}$ and $\varphi_{carr}$ are amplitude and phase of the output pulse carrier, $\varphi_{ref}$ is the phase of a reference sinusoidal signal.

Measurements of time gated insertion losses $IL_{TG}(x)$ and time-gated output phase $\varphi_{TG}(x)$ of the TIRUS were conducted by moving the specimen with respect to the sensor. The frequencies of the probing pulse carrier were 27 MHz, 20 MHz and 12 MHz, their durations were 10 μs. In Fig. 2 the averaged experimental results of $IL_{TG}$ and φ$_{TG}$ are presented. Here green lines $|EL(x)|$ and $EPh(x)$ show insertion losses and phase expected to be obtained in the experiment as a result of interaction of the sensor with a subsurface flaw.

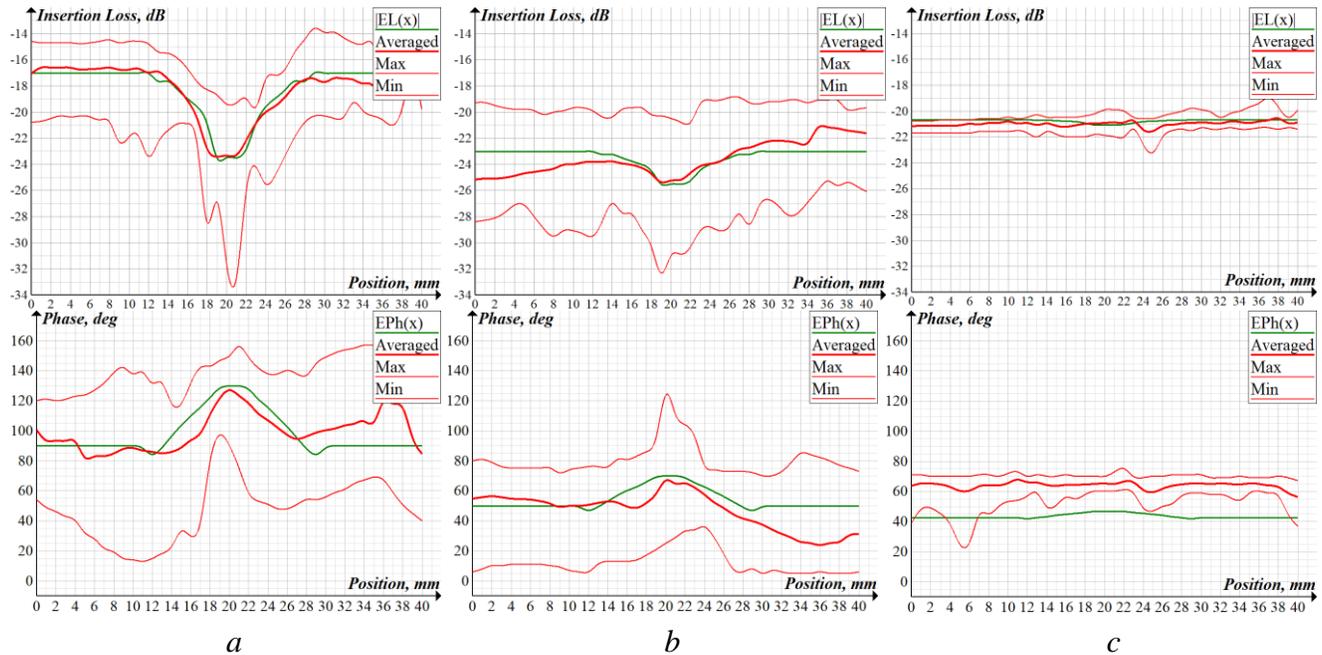

*Fig. 2. a – the averaged time-gated insertion loss IL$_{TG}$ and the averaged time-gated output phase φ$_{TG}$ of the TIRUS vs the sensor position x (0 to 40 mm) via the frequency 27 MHz of the probing pulse carrier; b – the averaged IL$_{TG}$ and the averaged φ$_{TG}$ of the TIRUS vs the sensor position x (0 to 40 mm) via the frequency 20 MHz of the probing pulse carrier; c – the averaged IL$_{TG}$ and the averaged φ$_{TG}$ of the TIRUS vs the sensor position x (0 to 40 mm) via the frequency 12 MHz of the probing pulse carrier; The MIN/MAX lines show the 95% confidence limits.*

The correlation coefficients between the expected and the averaged values are the following:

$$\rho_{IL, 27\ MHz} = 0,93; \quad \rho_{IL, 20\ MHz} = 0,47; \quad \rho_{IL, 12\ MHz} = 0,01;$$
$$\rho_{\varphi, 27\ MHz} = 0,56; \quad \rho_{\varphi, 20\ MHz} = 0,56; \quad \rho_{\varphi, 12\ MHz} = 0,12.$$

According to the results, measured insertion losses, in the area where there exists a subsurface defect, are up to 7 dB on the frequency 27 MHz of the probing pulse carrier, up to 3 dB on the frequency 20 MHz and up to 0,5 dB on the frequency 12 MHz. These results indicate a possibility of subsurface flaw detection via the Total Internal Reflection Ultrasonic Sensor using the ultrasonic frequencies band between 27 and 20 MHz.

## Conclusion

The TIRUS insertion losses and phase, that are its operational characteristics, were measured using time-gated (pulse) techniques. In the experiments, the TIRUS capability of sensing subsurface flaws has been proved on the ultrasonic frequencies of 27 and 20 MHz. Also, it has been established that experimental results correspond to the expected values on the frequencies band from 27 to 20 MHz.